\documentclass[prl,twocolumn,superscriptaddress, showpacs,preprintnumbers,amsmath,amssymb]{revtex4}

\usepackage{graphicx,color}
\usepackage{multirow}

\begin{document}

\preprint{}

\title{Finite bias Cooper pair splitting}

\author{L.~Hofstetter}
\affiliation{Department of Physics, University of Basel, Klingelbergstrasse 82, CH-4056 Basel, Switzerland}

\author{S.~Csonka}
\affiliation{Department of Physics, University of Basel, Klingelbergstrasse 82, CH-4056 Basel, Switzerland}
\affiliation{Department of Physics, Budapest University of Technology and Economics,\\ Budafoki u. 6, 1111 Budapest, Hungary}

\author{A.~Baumgartner}
\email{andreas.baumgartner@unibas.ch}
\affiliation{Department of Physics, University of Basel, Klingelbergstrasse 82, CH-4056 Basel, Switzerland}

\author{G.~F\"{u}l\"{o}p}
\affiliation{Department of Physics, Budapest University of Technology and Economics,\\ Budafoki u. 6, 1111 Budapest, Hungary}

\author{S.~d'Hollosy}
\affiliation{Department of Physics, University of Basel, Klingelbergstrasse 82, CH-4056 Basel, Switzerland}

\author{J.~Nyg{\aa}rd}
\affiliation{Nano-Science Center, Niels Bohr Institute, University of Copenhagen, Universitetsparken 5, DK-2100
Copenhagen, Denmark} 

\author{C.~Sch\"{o}nenberger}
\affiliation{Department of Physics, University of Basel, Klingelbergstrasse 82, CH-4056 Basel, Switzerland}

\date{\today}

\begin{abstract}
In a device with a superconductor coupled to two parallel quantum dots (QDs) the electrical tunability of the QD levels can be used to exploit non-classical current correlations due to the splitting of Cooper pairs. We experimentally investigate the effect of a finite potential difference across one quantum dot on the conductance through the other completely grounded QD in a Cooper pair splitter fabricated on an InAs nanowire. We demonstrate that the electrical transport through the device can be tuned by electrical means to be dominated either by Cooper pair splitting (CPS), or by elastic co-tunneling (EC). The basic experimental findings can be understood by considering the energy dependent density of states in a QD. The reported experiments add bias-dependent spectroscopy to the investigative tools necessary to develop CPS-based sources of entangled electrons in solid-state devices.
\end{abstract}


\maketitle

The electrons of a Cooper pair in a conventional superconductor form a spin singlet, which might be exploited as a naturally occurring on-chip source of spin-entangled Einstein-Podolsky-Rosen (EPR) \cite{Einstein_1935, Reid_RevModPhys81_2009} electron pairs, if the electrons can be separated coherently. This Cooper pair splitting (CPS) can be understood as inverse crossed Andreev reflection and was initially searched for in metallic nanostructures with tunnel contacts \cite{Beckmann_PRL93_2004, Russo_Klapwijk_PRL95_2005, Cadden-Zimansky_Chandrasekhar_NatPhys_2009, Kleine_Baumgartner_EPL87_2009, Wei_Chandrasekhar_NatPhys_2010}. However, other processes like elastic co-tunneling (EC) make the detection of CPS difficult. Electron-electron interactions are relevant in these processes, but can not be tuned by external means in these structures. It has been suggested to use electrically tunable quantum dots (QDs) coupled to a superconducting lead to obtain such tunability \cite{Recher_Loss_PRB63_2001, Sauret_2004_PRB70_2004}. Recently, first transport characteristics at zero bias were reported for InAs nanowire (NW) and carbon nanotube QD devices, which are explained by CPS \cite{Hofstetter2009, Herrmann_Kontos_Strunk_PRL104_2010}.

Bias-tunable non-local resistance in a metallic structure was attributed to the excitation of different modes of the electromagnetic environment by CPS and EC \cite{Russo_Klapwijk_PRL95_2005, Yeyati_Klapwijk_NaturePhys_2007, Kleine_Baumgartner_EPL87_2009}. This mechanism, however, is difficult to control experimentally. Here we report finite bias differential conductance measurements on InAs nanowire devices similar to those in Ref.~\cite{Hofstetter2009}. We show that the conductance through one QD is not entirely due to local processes, but also due to non-local higher order tunneling, consistent with a simple picture of CPS and EC. Our experiments establish an additional parameter and a procedure to characterize a generic Cooper pair splitter. Since the relevant processes depend differently on the bias, the relative rates can be tuned by external means. In particular, we show that the energy dependent effective density of states (DoS) due to the QD levels is crucial for the reported effects.

\begin{figure}[b]
\centering
\includegraphics{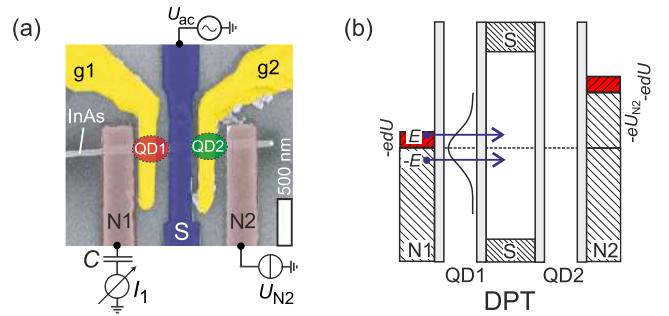}
\caption{(Color online) (a) Colored SEM image of a typical device and measurement circuitry. (b) Schematics of direct Cooper pair tunneling (DPT) with an ac voltage $dU$ imposed on QD1 and QD2 and a finite bias $U_{\rm N2}$ across QD2.}
\label{Figure1}
\end{figure}

A colored SEM image of a generic CPS device is shown in Fig.~1(a). An InAs wire \cite{Jespersen2006} is contacted by a superconducting strip in the center (S) and two Ti/Au normal metal leads (N1, N2), which define two quantum dots QD1 and QD2. The two top-gates (g1, g2) are strongly decoupled from each other, so that the QD levels can be tuned individually, while the highly-doped Si wafer separated from the device by $400\,$nm thermal oxide serves as a global back-gate. More details on the sample fabrication can be found in \cite{Hofstetter2009}. All experiments were performed at the base-temperature of $T\approx 20\,$mK.

Also shown in Fig.~1(a) is the measurement schematic: an ac voltage of $U_{\rm ac}=10\,\mu$V at $77\,$Hz is applied to the superconductor at zero dc potential. A dc voltage $U_{\rm N2}$ is applied to contact N2 of QD2, while a home-built current-voltage converter (gain $10^8$\,V/A) and a standard lock-in amplifier are used to measure the ac current $I_1$ in contact N1 from QD1. A $10\,\mu$F capacitor $C$ is used to decouple possible IV-converter off-set voltages from the device, while keeping the dc potential at zero. Figure~1(b) shows the energy diagram of such a device.

\begin{figure}[t]
\centering
\includegraphics[width=0.5\textwidth]{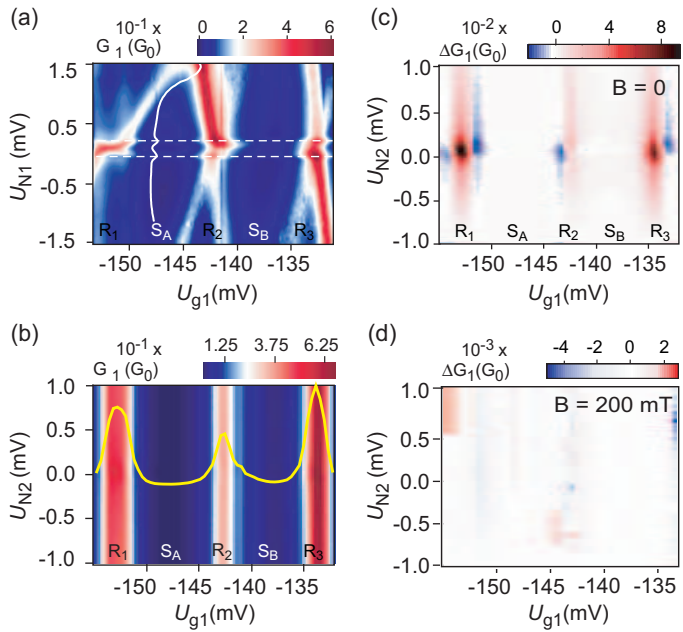}
\caption{(Color online) (a) Characterization of QD1: differential conductance $G_1$ as a function of $U_{\rm N1}$ and $U_{\rm g1}$ at $B = 0$. The cross section shows two small features related to the superconductor gap $\Delta$. (b) $G_1$ as a function of $U_{\rm N2}$ and $U_{\rm g1}$ at $B = 0$. (c) $\Delta G_1=G_1(U_{\rm g1},U_{\rm N2})-G_1(U_{\rm g1},U_{\rm N2}=-1\,\rm{mV})$ derived from the data in (b) for the same parameter range. (d) $\Delta G_1$ for the same experiment at $B = 200\,$mT.}
\label{Figure2}
\end{figure}

To characterize QD1, the differential conductance $G_1=dI_1/dU$ is plotted in Fig.~2(a) as a function of the gate voltage $U_{\rm g1}$ and the bias applied to N1, $U_{\rm N1}$. We observe well-defined Coulomb blockade oscillations consistent with the formation of a QD \cite{Csonka2008}. We label the resonances in the figure as R$_1$-R$_3$ and the electronic states in-between as S$_{\rm A}$ and S$_{\rm B}$, from left to right. From this plot we extract various sample parameters, e.g. the superconducting gap $\Delta \approx 130~\mu$eV (cross section at $U_{\rm g1}=-0.146\,$V), or an addition energy of $\sim3\,$meV and the tunnel coupling of $\Gamma \approx 500~\mu eV$ for S$_{\rm B}$. In addition, a Kondo resonance \cite{Goldhaber-Gordon1998} in S$_{\rm B}$ can be observed at zero bias in the normal state data (not shown). A similar characterization of QD2 in the parameter range of the experiments (not shown) exhibits only small variations in the conductance. In the discussion below we therefore assume a resonance in the effective DoS of QD1 and a constant DoS for QD2, as illustrated in Fig.~1(b).

The dominant component of $I_1$ is due to local transport processes between $S$ and $N_1$. As an example, direct Cooper pair tunneling (DPT), also known as Andreev reflection, is illustrated in Fig.~1(b): applying the voltage $dU$ to S as in the experiment is equivalent to applying $-dU$ to N1 and N2. An electron at energy $E$ and one at $-E$ are injected from N1 to form a Cooper pair in S. Because $S$ and $N_1$ are at the same dc potential, all local transport processes are independent of the voltage $U_{\rm N2}$ on N2, if charge imbalance \cite{Kleine_Baumgartmer_Nanotechnology21_2010} and local heating can be neglected. Therefore only 'non-local' transport processes, which comprise correlated coherent tunneling of electrons through QD1 {\it and} QD2, can lead to a dependence of $I_1$ on $U_{\rm N2}$.

Figure~2(b) shows $G_1$ as a function of the top gate voltage $U_{\rm g1}$ of QD1 and the bias $U_{\rm N2}$ applied to QD2. On this scale the conductance seems independent of $U_{\rm N2}$.
We assume that all non-local processes become ineffective at $U_{\rm N2}\gg\Delta/e$ and therefore plot in Fig.~2(c) the deviation of $G_1$ from a high-bias value, $\Delta G_1(U_{\rm g1}, U_{\rm N2})=G_1(U_{\rm g1}, U_{\rm N2})-G_1(U_{\rm g1}, U_{\rm N2}=-1\,{\rm mV})$.
We note that the colorscale is adjusted to white representing $\Delta G_1=0$. For the gate voltage $U_{\rm g1}$ tuned to resonance R$_1$, $\Delta G_1$ has a positive maximum at zero bias with $\Delta G_1/G_1\approx 15\%$. The extent of this maximum is roughly $eU_{\rm N2}\approx\Delta$ and $e\alpha U_{\rm g1}\approx \Gamma$ ($\alpha$: lever arm of top gate g1). With $U_{\rm g1}$ slightly off-peak, a minimum in $\Delta G_1$ occurs on both sides of resonance R$_1$, but {\it not} centered around $U_{\rm N2}=0$. These features will be examined in more detail below. Similar features occur on the resonances 2 and 3, however without minima where state B is involved.

In a control experiment the superconductivity is suppressed by an external magnetic field of $200\,$mT parallel to the Al strip. The corresponding plot of $\Delta G_1$ is shown in Fig.~2(d). No clear structure can be discriminated, though the scale is considerably smaller than in Fig.~2(c). We therefore conclude that the features in Fig.~2(c) are due to the superconductor and non-local transport processes, which become unlikely if S is in the normal-conducting state.

\begin{figure}[t]
\centering
\includegraphics[width=0.5\textwidth]{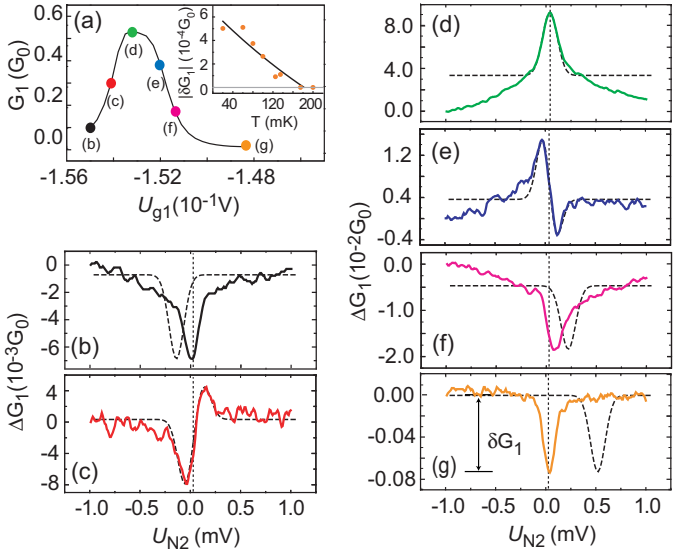}
\caption{(Color online) Bias dependence of $\Delta G_1$ for a series of top gate voltages $U_{\rm g1}$. The latter are indicated in (a). The dashed curves are derived from the model discussed in the text. Vertical lines indicate $U_{\rm N2}=0$, including a small offset from the IV-converter on N2. The inset in (a) shows the temperature dependence of the the minimum in (g) with a black line as a guide to the eye.}
\label{Figure3}
\end{figure}

Figure~3 shows cross sections $\Delta G_1(U_{\rm N2})$ of Fig.~2(c). The corresponding top-gate voltages $U_{\rm g1}$ are near resonance 1 and indicated on the zero bias top-gate sweep $G_1(U_{\rm g1})$ in Fig.~3(a). For gate voltages far from the resonance the bias dependence of $\Delta G_1$ shows a small but pronounced minimum at zero bias, as shown in Figs.~3(b) and (g), while on resonance a strong peak is observed at zero bias, see Fig.~3(d). Zero bias across QD2 is indicated by dashed vertical lines in all plots and is slightly offset due to a small input offset-voltage from an IV-converter mounted on lead N2. Slightly off-resonance the bias dependence is asymmetric with respect to zero bias: in Fig.~3(c) $\Delta G_1$ exhibits a minimum at negative and a maximum at positive bias, while on the other side of the resonance a maximum can be found at negative and a minimum at positive bias, see Figs.~3(e) and (f). We note that the maxima and minima near the resonance are not at zero bias, whereas far off the resonance only a minimum at $U_{\rm N2}=0$ is found.

The non-local signals decay with increasing temperature, as shown in the inset of Fig.~3(a) for the conductance minimum  at $U_{\rm g1}=-0.1485\,$V, essentially the gate position of Fig.~3(g). The amplitude of $\Delta G_1$ decreases monotonically and disappears at $T \approx 175\,$mK. Up to this temperature we have found no significant change in the superconductor gap. We therefore conclude that $\Delta$ is not the limiting energy scale in this problem, reminiscent of the CPS observed in \cite{Hofstetter2009} at zero-bias.


A qualitative understanding of the experiments can be gained by expressing the CPS and EC tunneling rates by energy dependent effective density of states $D_1(E)$ and $D_2(E)$ in QD1 and QD2, incorporating the respective QD transmissions. The processes are illustrated in Fig.~4, for which we assume a resonance feature for $D_1$ and a constant for QD2, i.e. $D_2(E)=D_2$. We neglect changes in the DoS by virtual tunneling processes \cite{Chevallier_Martin_PRB83_2011} and electron-electron interactions in the QDs and the leads.

\begin{figure}[b]
\centering
\includegraphics{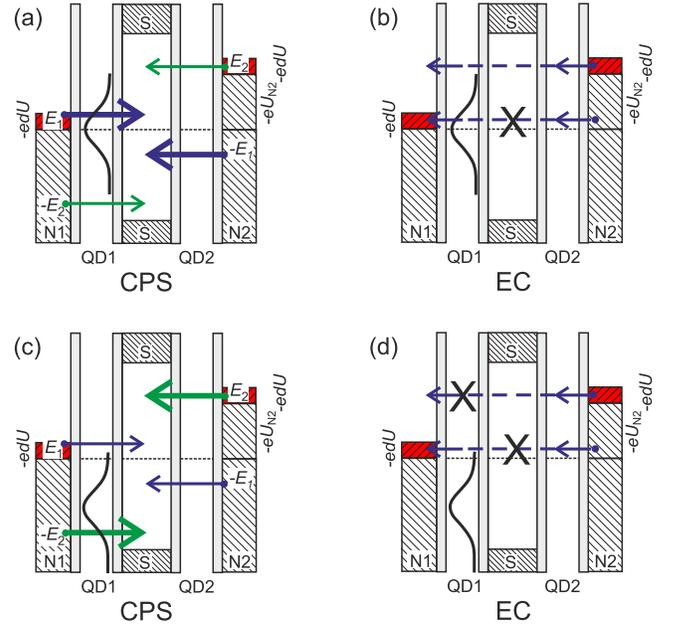}
\caption{(Color online) Schematics of CPS and EC at finite bias on QD2 and an ac voltage $dU$ imposed on both QDs. (a) and (b) show the situation when QD1 is tuned to a resonance, while (c) and (d) depict an off-resonance case.}
\label{Figure4}
\end{figure}

The probability that an electron from QD1 at energy $E$ and one from QD2 at energy $-E$ form a Cooper pair in S (inverse CPS) depends on the DoS as
\begin{equation}
\propto M_{\rm CPS}(E)D_1(E)D_2(-E)f(E,\mu_1)f(-E,\mu_2),
\end{equation}
with the respective electrochemical potentials $\mu_{1,2}$ in the leads, measured relative to the $\mu_{\rm S}=0$ in the superconductor. $M_{\rm CPS}$ represents the transition probability. For CPS the states of the QDs have to be empty initially and the Fermi functions $f(\pm E,\mu_{1,2})$ are replaced by $1-f(\pm E,\mu_{1,2})$. We note that generally we have inverse CPS for $U_{\rm N2}<0$ and CPS for $U_{\rm N2}>0$. The situation for EC is shown schematically in Fig.~4(b). Since the process is elastic, its probability at the energy $E$ scales as
\begin{equation}
\propto M_{\rm EC}(E)D_1(E)D_2(E)\left[f(E,\mu_2)-f(E,\mu_1)\right].
\end{equation}
Summation over all energies for periodically varying $U_{\rm N1}$ and $U_{\rm N2}$ results in the non-local (nl) differential conductance, $\Delta G_1=dI_{1,{\rm nl}}/dU$. Assuming constant $M_{\rm CPS}$, $M_{\rm EC}$ and $D_2$, as well as $T\rightarrow 0$ and $edU<<\Gamma$, the following intuitive expressions for the contributions of CPS ($G_{\rm CPS}$) and EC ($G_{\rm EC}$) to $\Delta G_1 = G_{\rm CPS} + G_{\rm EC}$ can be found:
\begin{eqnarray}
G_{\rm CPS}&=& \frac{e^2}{h}M_{\rm CPS}D_2\left[ D_1(0) +D_1(+eU_{\rm N2})\right]
\label{model_eq_1} \\
G_{\rm EC}&=& \frac{e^2}{h}M_{\rm EC}D_2\left[ D_1(0) -D_1(-eU_{\rm N2})\right]
\label{model_eq_2}
\end{eqnarray}
Both processes have a component from the DoS at $E=0$. In addition, CPS has a positive contribution from $D_1$ at $+eU$, while EC has a negative contribution from $-eU$. This can also be understood from the diagrams in Fig.~4.
We note that in our measurement scheme the differential conductance due to EC would be zero for energy independent transmissions, e.g. in metallic structures, and the contribution of CPS would be constant and independent of the bias applied to N2.

First we discuss the case when QD1 is tuned to a symmetric resonance, i.e. $D_1(E)=D_1(-E)$ and $D_1(0)>D_1(E\neq 0)$, as depicted in Fig.~4(a) and (b). From Eqs.~(\ref{model_eq_1}) and (\ref{model_eq_2}) one finds that $G_{\rm CPS}$ has a maximum and $G_{\rm EC}$ a minimum with $G_{\rm EC}=0$ at $U_{\rm N2}=0$. Both signals are positive on a resonance. If the two processes had the same probability, $M_{\rm CPS}=M_{\rm EC}$, the variations of $G_{\rm CPS}$ and $G_{\rm EC}$ would cancel exactly and we would not expect any changes in $\Delta G_1$ as a function of $U_{\rm N2}$. We therefore conclude that CPS and EC can obtain different relative weights and the comparison with the experiment in Fig.~3(d) suggests that on a resonance CPS is the dominant process, i.e. $M_{\rm CPS}\gg M_{\rm EC}$, independent of $U_{\rm N2}$.

An off-resonance situation is illustrated in Figs.~4(c) and (d). If the resonance is shifted in energy to $E=E_0=-e\alpha\Delta U_{\rm g1}$ by top-gate 1, Eqs.~(\ref{model_eq_1}) and (\ref{model_eq_2}) predict two features: 1) Since $D_1(E_0) \gg D_1(0) > D_1(-E_0)$, one finds for $eU_{\rm N2}=E_0$ that $G_{\rm CPS}$ has a positive maximum and dominates $\Delta G_1$. 2) Similarly, one finds for $eU_{\rm N2}=-E_0$ a negative minimum in both, $G_{\rm EC}$ and $\Delta G_1$, the latter because $G_{\rm CPS}$ is negligible. Such features are observed in Fig.~3(c) and (e), where a minimum and a maximum occur as a function of $U_{\rm N2}$, with reversed order depending on the gate voltage relative to the resonance.

The results of numerical evaluations of Eqs.~(2) and (3) with a Gaussian for $D_1$ and the adjustable parameters $M_{\rm CPS}$, $M_{\rm EC}$ and an offset, leads to the dashed curves in Figs.~3(b)-(g). The data are reproduced qualitatively near the resonance and we find that CPS is favored on resonance, while slightly off-resonance EC obtains a similar strength as CPS. These findings might be related to a reduction of CPS for energies close to $\Delta$, while EC is increased due to the energy-dependence of the second order tunnel matrix elements. Both rates diminish above $\Delta$ due to the creation of excitations in S.

This model is very rudimentary and neglects electron-electron interactions in the leads and on the QDs, which might be crucial to understand details in our data. A strong indication for this is that in all Coulomb blockade regions a minimum occurs at zero bias, see Figs.~3(b) and (g), which can not be reproduced in the model. The dashed curves in these figures are for $M_{\rm EC}\gg M_{\rm CPS}$ and show a minimum adjusted to the dip in the data, but at a wrong bias. A dip at zero bias, however, is in qualitative agreement with a zero-bias anomaly due to dynamical Coulomb blockade at low transmissions \cite{Kleine_Baumgartner_EPL87_2009}. In addition, in state S$_{\rm B}$, for which we observe Kondo correlations in the Coulomb diamonds, the minima in $\Delta G_1$ expected near the resonances are missing or very weak, while they are clearly observed for the other states. This might be due to a competition between the superconducting and the Kondo correlations, which is not accounted for in our model. 

In summary, we have reported finite bias measurements on an InAs nanowire quantum dot Cooper pair splitter. Our results show that the energy dependent transmission due to the QDs has strong effects on non-local processes in such systems. We show that finite-bias spectroscopy is useful to identify non-local processes and find that Cooper pair splitting can be the dominant process on a QD resonance. Off-resonance we find a pattern consistent with EC being of similar strength as CPS. These findings are interpreted in a simple model based on energy dependent effective density of states. However, this model does not account for the relative strength of the processes, nor for features in the Coulomb blockade regions or where Kondo correlations are relevant. We tentatively attribute the latter findings to electron-electron interactions, which should be studied in greater detail to understand the involved mechanisms and find means to exploit them in an on-chip source of entangled electrons.

We thank Jens Schindele for fruitful discussions and gratefully acknowledge the financial support by the EU FP7 project SE$^2$ND, the EU ERC project CooPairEnt, the Swiss NCCR Nano and NCCR Quantum, the Swiss SNF, and the Danish Research Councils.

\bibliographystyle{apsrev}

\begin{thebibliography}{99}
\expandafter\ifx\csname natexlab\endcsname\relax\def\natexlab#1{#1}\fi
\expandafter\ifx\csname bibnamefont\endcsname\relax
  \def\bibnamefont#1{#1}\fi
\expandafter\ifx\csname bibfnamefont\endcsname\relax
  \def\bibfnamefont#1{#1}\fi
\expandafter\ifx\csname citenamefont\endcsname\relax
  \def\citenamefont#1{#1}\fi
\expandafter\ifx\csname url\endcsname\relax
  \def\url#1{\texttt{#1}}\fi
\expandafter\ifx\csname urlprefix\endcsname\relax\def\urlprefix{URL }\fi
\providecommand{\bibinfo}[2]{#2}
\providecommand{\eprint}[2][]{\url{#2}}


\bibitem{Einstein_1935}
\bibinfo{author}{\bibfnamefont{A.}~\bibnamefont{Einstein}},
\bibinfo{author}{\bibfnamefont{B.}~\bibnamefont{Podolsky}},
\bibnamefont{and}
\bibinfo{author}{\bibfnamefont{N.}~\bibnamefont{Rosen}},
\bibinfo{journal}{Phys. Rev.}
\textbf{\bibinfo{volume}{47}},
\bibinfo{pages}{0777}
(\bibinfo{year}{1935}).

\bibitem{Reid_RevModPhys81_2009}
\bibinfo{author}{\bibfnamefont{M.D.}~\bibnamefont{Reid}},
\bibinfo{author}{\bibfnamefont{P.D.}~\bibnamefont{Drummond}},
\bibinfo{author}{\bibfnamefont{W.P.}~\bibnamefont{Bowen}},
\bibinfo{author}{\bibfnamefont{E.G.}~\bibnamefont{Cavalcanti}},
\bibinfo{author}{\bibfnamefont{P.K.}~\bibnamefont{Lam}},
\bibinfo{author}{\bibfnamefont{H.A.}~\bibnamefont{Bachor}},
\bibinfo{author}{\bibfnamefont{U.L.}~\bibnamefont{Andersen}},
\bibnamefont{and}
\bibinfo{author}{\bibfnamefont{G.}~\bibnamefont{Leuchs}},
\bibinfo{journal}{Rev. Mod. Phys.}
\textbf{\bibinfo{volume}{81}},
\bibinfo{pages}{1727}
(\bibinfo{year}{2009}).

\bibitem{Beckmann_PRL93_2004}
\bibinfo{author}{\bibfnamefont{D.}~\bibnamefont{Beckmann}},
\bibinfo{author}{\bibfnamefont{H.B.}~\bibnamefont{Weber}},
\bibnamefont{and}
\bibinfo{author}{\bibfnamefont{H.}~\bibnamefont{v. L\"ohneysen}},
\bibinfo{journal}{Phys. Rev. Lett.}
\textbf{\bibinfo{volume}{93}},
\bibinfo{pages}{197003}
(\bibinfo{year}{2004}).

\bibitem{Russo_Klapwijk_PRL95_2005}
\bibinfo{author}{\bibfnamefont{S.}~\bibnamefont{Russo}},
\bibinfo{author}{\bibfnamefont{M.}~\bibnamefont{Kroug}},
\bibinfo{author}{\bibfnamefont{T.M.}~\bibnamefont{Klapwijk}},
\bibnamefont{and}
\bibinfo{author}{\bibfnamefont{A.F.}~\bibnamefont{Morpurgo}},
\bibinfo{journal}{Phys. Rev. Lett.}
\textbf{\bibinfo{volume}{95}},
\bibinfo{pages}{027002}
(\bibinfo{year}{2005}).

\bibitem{Cadden-Zimansky_Chandrasekhar_NatPhys_2009}
\bibinfo{author}{\bibfnamefont{P.}~\bibnamefont{Cadden-Zimansky}},
\bibinfo{author}{\bibfnamefont{J.}~\bibnamefont{Wei}},
\bibnamefont{and}
\bibinfo{author}{\bibfnamefont{V.}~\bibnamefont{Chandrasekhar}},
\bibinfo{journal}{Nature Phys.}
\textbf{\bibinfo{volume}{5}},
\bibinfo{pages}{393}
(\bibinfo{year}{2009}).

\bibitem{Kleine_Baumgartner_EPL87_2009}
\bibinfo{author}{\bibfnamefont{A.}~\bibnamefont{Kleine}},
\bibinfo{author}{\bibfnamefont{A.}~\bibnamefont{Baumgartner}},
\bibinfo{author}{\bibfnamefont{J.}~\bibnamefont{Trbovic}},
\bibnamefont{and}
\bibinfo{author}{\bibfnamefont{C.}~\bibnamefont{Sch\"onenberger}},
\bibinfo{journal}{Europhys. Lett.}
\textbf{\bibinfo{volume}{87}},
\bibinfo{pages}{27011}
(\bibinfo{year}{2009}).

\bibitem{Wei_Chandrasekhar_NatPhys_2010}
\bibinfo{author}{\bibfnamefont{J.}~\bibnamefont{Wei}},
\bibnamefont{and}
\bibinfo{author}{\bibfnamefont{V.}~\bibnamefont{Chandrasekhar}},
\bibinfo{journal}{Nature Phys.}
\textbf{\bibinfo{volume}{6}},
\bibinfo{pages}{494}
(\bibinfo{year}{2010}).

\bibitem{Recher_Loss_PRB63_2001}
\bibinfo{author}{\bibfnamefont{P.}~\bibnamefont{Recher}},
\bibinfo{author}{\bibfnamefont{E.V.}~\bibnamefont{Sukhorukov}},
\bibnamefont{and}
\bibinfo{author}{\bibfnamefont{D.}~\bibnamefont{Loss}},
\bibinfo{journal}{Phys. Rev. B}
\textbf{\bibinfo{volume}{63}},
\bibinfo{pages}{165314}
(\bibinfo{year}{2001}).

\bibitem{Sauret_2004_PRB70_2004}
\bibinfo{author}{\bibfnamefont{O.}~\bibnamefont{Sauret}},
\bibinfo{author}{\bibfnamefont{D.}~\bibnamefont{Feinberg}},
\bibnamefont{and}
\bibinfo{author}{\bibfnamefont{T.}~\bibnamefont{Martin}},
\bibinfo{journal}{Phys. Rev. B}
\textbf{\bibinfo{volume}{70}},
\bibinfo{pages}{245313}
(\bibinfo{year}{2004}).

\bibitem{Hofstetter2009}
\bibinfo{author}{\bibfnamefont{L.}~\bibnamefont{Hofstetter}},
\bibinfo{author}{\bibfnamefont{S.}~\bibnamefont{Csonka}},
\bibinfo{author}{\bibfnamefont{J.}~\bibnamefont{Nyg{\aa}rd}},
\bibnamefont{and}
\bibinfo{author}{\bibfnamefont{C.}~\bibnamefont{Sch\"{o}nenberger}},
\bibinfo{journal}{Nature} \textbf{\bibinfo{volume}{461}},
\bibinfo{pages}{960} (\bibinfo{year}{2009}).

\bibitem{Herrmann_Kontos_Strunk_PRL104_2010}
\bibinfo{author}{\bibfnamefont{L.G.}~\bibnamefont{Herrmann}},
\bibinfo{author}{\bibfnamefont{F.}~\bibnamefont{Portier}},
\bibinfo{author}{\bibfnamefont{P.}~\bibnamefont{Roche}},
\bibinfo{author}{\bibfnamefont{A.}~\bibnamefont{Levy Yeyati}},
\bibinfo{author}{\bibfnamefont{T.}~\bibnamefont{Kontos}},
\bibnamefont{and}
\bibinfo{author}{\bibfnamefont{C.}~\bibnamefont{Strunk}},
\bibinfo{journal}{Phys. Rev. Lett.} \textbf{\bibinfo{volume}{104}},
\bibinfo{pages}{026801} (\bibinfo{year}{2010}).

\bibitem{Yeyati_Klapwijk_NaturePhys_2007}
\bibinfo{author}{\bibfnamefont{A.}~\bibnamefont{Levy Yeyati}},
\bibinfo{author}{\bibfnamefont{F.S.}~\bibnamefont{Bergeret}},
\bibinfo{author}{\bibfnamefont{A.}~\bibnamefont{Mart\'{i}n-Rodero}},
\bibnamefont{and}
\bibinfo{author}{\bibfnamefont{T.M.}~\bibnamefont{Klapwijk}},
\bibinfo{journal}{Nature Phys.} \textbf{\bibinfo{volume}{3}},
\bibinfo{pages}{455} (\bibinfo{year}{2007}).

\bibitem{Jespersen2006}
\bibinfo{author}{\bibfnamefont{T.~S.} \bibnamefont{Jespersen}},
\bibinfo{author}{\bibfnamefont{M.}~\bibnamefont{Aagesen}},
\bibinfo{author}{\bibfnamefont{C.}~\bibnamefont{Sorensen}},
\bibinfo{author}{\bibfnamefont{P.~E.} \bibnamefont{Lindelof}},
\bibnamefont{and} \bibinfo{author}{\bibfnamefont{J.}~\bibnamefont{Nyg{\aa}rd}},
\bibinfo{journal}{Phys. Rev. B} \textbf{\bibinfo{volume}{74}},
\bibinfo{pages}{233304} (\bibinfo{year}{2006}).

\bibitem{Csonka2008}
\bibinfo{author}{\bibfnamefont{S.}~\bibnamefont{Csonka}},
\bibinfo{author}{\bibfnamefont{L.}~\bibnamefont{Hofstetter}},
\bibinfo{author}{\bibfnamefont{F.}~\bibnamefont{Freitag}},
\bibinfo{author}{\bibfnamefont{S.}~\bibnamefont{Oberholzer}},
\bibinfo{author}{\bibfnamefont{T.~S.} \bibnamefont{Jespersen}},
\bibinfo{author}{\bibfnamefont{M.}~\bibnamefont{Aagesen}},
\bibinfo{author}{\bibfnamefont{J.}~\bibnamefont{Nyg{\aa}rd}},
\bibnamefont{and}
\bibinfo{author}{\bibfnamefont{C.}~\bibnamefont{Sch\"{o}nenberger}},
\bibinfo{journal}{Nano Lett.} \textbf{\bibinfo{volume}{8}},
\bibinfo{pages}{3932} (\bibinfo{year}{2008}).

\bibitem{Goldhaber-Gordon1998}
\bibinfo{author}{\bibfnamefont{D.}~\bibnamefont{Goldhaber-Gordon}},
\bibinfo{author}{\bibfnamefont{H.}~\bibnamefont{Shtrikman}},
\bibinfo{author}{\bibfnamefont{D.}~\bibnamefont{Mahalu}},
\bibinfo{author}{\bibfnamefont{D.}~\bibnamefont{Abusch-Magder}},
\bibinfo{author}{\bibfnamefont{U.}~\bibnamefont{Meirav}}, \bibnamefont{and}
\bibinfo{author}{\bibfnamefont{M.~A.} \bibnamefont{Kastner}},
\bibinfo{journal}{Nature} \textbf{\bibinfo{volume}{391}},
\bibinfo{pages}{156} (\bibinfo{year}{1998}).

\bibitem{Kleine_Baumgartmer_Nanotechnology21_2010}
\bibinfo{author}{\bibfnamefont{A.}~\bibnamefont{Kleine}},
\bibinfo{author}{\bibfnamefont{A.}~\bibnamefont{Baumgartner}},
\bibinfo{author}{\bibfnamefont{J.}~\bibnamefont{Trbovic}},
\bibinfo{author}{\bibfnamefont{D.S.}~\bibnamefont{Golubev}},
\bibinfo{author}{\bibfnamefont{A.D.}~\bibnamefont{Zaikin}},
\bibnamefont{and}
\bibinfo{author}{\bibfnamefont{C.}~\bibnamefont{Sch\"{o}nenberger}},
\bibinfo{journal}{Nanotechnology} \textbf{\bibinfo{volume}{21}},
\bibinfo{pages}{274002} (\bibinfo{year}{2010}).

\bibitem{Falci_EPL54_2001}
\bibinfo{author}{\bibfnamefont{G.}~\bibnamefont{Falci}},
\bibinfo{author}{\bibfnamefont{D.}~\bibnamefont{Feinberg}},
\bibnamefont{and}
\bibinfo{author}{\bibfnamefont{F.W.J.}~\bibnamefont{Hekking}},
\bibinfo{journal}{Europhys. Lett.}
\textbf{\bibinfo{volume}{54}},
\bibinfo{pages}{255}
(\bibinfo{year}{2001}).

\bibitem{Chevallier_Martin_PRB83_2011}
\bibinfo{author}{\bibfnamefont{D.}~\bibnamefont{Chevallier}},
\bibinfo{author}{\bibfnamefont{J.}~\bibnamefont{Rech}},
\bibinfo{author}{\bibfnamefont{T.}~\bibnamefont{Jonckheere}},
\bibnamefont{and}
\bibinfo{author}{\bibfnamefont{T.}~\bibnamefont{Martin}},
\bibinfo{journal}{Phys. Rev. B}
\textbf{\bibinfo{volume}{83}},
\bibinfo{pages}{125421}
(\bibinfo{year}{2011}).

\end{thebibliography}

\end{document}